\documentclass[amsmath,reprint,amssymb,prl, aps, longbibliography]{revtex4-2} 
\usepackage{graphicx}
\usepackage{dcolumn}
\usepackage{bm}
\usepackage[colorlinks,citecolor=blue,linkcolor=blue]{hyperref} 
\usepackage{color}  

\usepackage{ulem}

\begin{document}

\preprint{AIP/123-QED}

\title{Always-Real-Eigenvalued Non-Hermitian Topological Systems}

\author{Yang Long}
\author{Haoran Xue}
\author{Baile Zhang}
\email{Corresponding author: blzhang@ntu.edu.sg}
\affiliation{%
Division of Physics and Applied Physics, School of Physical and Mathematical Sciences, Nanyang Technological University, 21 Nanyang Link, Singapore 637371, Singapore
}%

\date{\today}

\begin{abstract}
The effect of non-Hermiticity in band topology has sparked many discussions on non-Hermitian topological physics. It has long been known that non-Hermitian Hamiltonians can exhibit real energy spectra under the condition of parity-time ($PT$) symmetry---commonly implemented with balanced loss and gain---but only when non-Hermiticity is relatively weak. Sufficiently strong non-Hermiticity, on the other hand, will destroy the reality of energy spectra, a situation known as spontaneous $PT$-symmetry breaking. Here, based on non-reciprocal coupling, we show a systematic strategy to construct non-Hermitian topological systems exhibiting bulk and boundary energy spectra that are always real, regardless of weak or strong non-Hermiticity. Such nonreciprocal-coupling-based non-Hermiticity can directly drive a topological phase transition and determine the band topology, as demonstrated in a few non-Hermitian systems from 1D to 2D. Our work develops so far the only theory that can guarantee the reality of energy spectra for non-Hermitian Hamiltonians, and offers a new avenue to explore non-Hermitian topological physics.
\end{abstract}
                    
\maketitle
Conservation of energy in quantum physics demands real eigenenergies for a closed system that is described by a Hermitian Hamiltonian. 
In presence of energy exchange with the surrounding environment, the energy conservation is broken. In such a situation, a non-Hermitian Hamiltonian will arise, leading to complex eigenvalues. 
However, the non-Hermiticity is not a sufficient condition for the existence of complex spectra, which means that it is possible to find a non-Hermitian system with real eigenvalues.

Parity-time($PT$) symmetry, one of the major discoveries in non-Hermitian quantum physics, claims that a class of non-Hermitian Hamiltonians with $PT$-symmetry can still have real spectra~\cite{Bender1998, ElGanainy2018}. 
This counterintuitive discovery  fundamentally overturned the past perception that only Hermitian operators can have real eigenvalues.
However, the $PT$ symmetry approach has a long-lasting limitation: it takes effect only when the non-Hermiticity is relatively weak ~\cite{Leykam2017,  Takata2018, ElGanainy2018, Song2020, Ao2020}; in other words, sufficiently strong non-Hermiticity will destroy the reality of energy spectra even when $PT$ symmetry is still respected — a situation known as spontaneous $PT$-symmetry breaking ~\cite{Bender1999, Oezdemir2019, ElGanainy2018, Miri2019, Feng2017,Weimann2016}. Therefore, the $PT$ symmetry cannot guarantee the reality of energy spectra. 
A $PT$-symmetric Hamiltonian belongs to a more general class of non-Hermitian Hamiltonians known as pseudo-Hermitian Hamiltonians~\cite{Mostafazadeh2002, Zhang2020a}, but pseudo-Hermiticity does not guarantee the reality of energy spectra either. For example, let us consider $H = i\sigma_z$, which is pseudo-Hermitian since $H^\dagger = \eta H \eta^{-1}$, where $\eta = \sigma_x$, but $H$ has complex  eigenvalues of $\pm i$. 
To our knowledge, there is no theory so far that can guarantee the reality of energy spectra in non-Hermitian systems.

Recently, there have been a lot of efforts in constructing topological states in non-Hermitian systems~\cite{Bergholtz2021, Song2019a, Weidemann2020, zhang2021acoustic}. 
This combination of topological physics and non-Hermitian physics challenges the conventional understanding of topological phases~\cite{Lee2016, Gong2018,Kawabata2019a,
Yao2018,Yao2018a,Gong2018,
Song2019,Yi2020,Okuma2020,
Zhang2020,Borgnia2020,Yoshida2020, Yoshida2019, Isobe2021}, as they were previously defined and classified based on their Hermitian Hamiltonians~\cite{Xiao2010, Hasan2010}. 
As a result, there have been many developments for non-Hermitian topological invariants to characterize non-Hermitian topology~\cite{Yao2018,Yao2018a, Leykam2017,Shen2018,Song2019,Yokomizo2019}. 
Another challenge that is still under active exploration is whether the non-Hermitian topologial systems can exhibit real energy spectra. 
Most studies along this line have adopted the $PT$ symmetry in order to maintain the real spectra as much as possible~\cite{Hu2011, Rudner2009, Esaki2011, Sato2012}. 
However, all these systems will necessarily exhibit complex spectra in presence of sufficiently strong non-Hermiticity. 
Moreover, the nontrivial topology in non-Hermitian topological systems is not necessarily determined by Hermitian parameters~\cite{Yao2018,Kawabata2020, Ghorashi2021, Ghorashi2021a,Yao2018a}, but can also be induced directly by non-Hermiticity. 
For example, recent studies have shown that by introducing deliberately designed loss and gain, a topological phase transition can be induced to generate topological states in non-Hermitian systems~\cite{Weimann2016, Takata2018, Xue2020, Gao2021, Luo2019}. 
Non-reciprocal coupling is another form of non-Hermiticity, but discussions on its directly induced topological phase transition [e.g., to induce a one-dimensional (1D) nontrivial Zak phase] have been relatively few.


In this Letter, we develop an approach to construct non-Hermitian topological systems whose energy spectra are always real, regardless of weak or strong non-Hermiticity. Due to the always real energy spectra, the constructed non-Hermitian systems will not exhibit winding topology in the complex energy plane~\cite{Bergholtz2021,  Kawabata2019a}, as in the recently discovered skin effect~\cite{Lee2016, Yao2018,Gong2018}, but will have nontrivial topology defined in momentum space. In particular, the nontrivial topology can be either inherited from the Hermitian counterpart, or induced by non-Hermiticity directly. The approach is developed by the non-commutative matrix production between a non-uniform diagonal matrix and a Hermitian Hamiltonian matrix. The resultant non-Hermitian systems can be implemented with non-reciprocal coupling.
We demonstrate the effectiveness of this approach in several concrete examples. 
In a 1D non-Hermitian system, we show that the band topology characterized by Zak phase can be inherited from the Hermitian Su-Schrieffer-Heeger (SSH) model, or induced by the non-Hermiticity.
Starting with a two-dimensional (2D) $C_3$-symmetric Hermitian system that is topologically trivial, we show that the non-zero topological polarization and fractional charges can emerge after introducing non-Hermitian terms, leading to topological corner states.
All these examples exhibit real enerey spectra for both bulk and boundary states, even in presence of strong non-Hermiticity.

Let us consider the following equation, which is commonly studied in the regular Sturm–Liouville theory~\cite{zettl2010sturm}:
\begin{equation}
H_0 \Psi_n = E_n M \Psi_n,
\label{eq:dynamics}
\end{equation}
where $H_0$ is a Hermitian matrix, $M$ is a real diagonal matrix with diagonal elements  $M_{ii} > 0$ and $\Psi_n$ is the column vector. 
One can obtain the eigenvalues of Eq.~\ref{eq:dynamics} by solving $|H_0-E_n M|=0$ and $E_n$ will be real ($E_n \in \mathbb{R}$) according to the regular Sturm–Liouville theory~\cite{zettl2010sturm}. 
Here, we consider $M = {\rm diag}[\varepsilon_1, \varepsilon_2, \varepsilon_3 ...\varepsilon_N]$ with $\varepsilon_i > 0$, and its determinant is nonzero (\textit{i.e}, $|M|\neq 0$) so that $M^{-1}$ exists. Based on these facts, we can construct a system with the Hamiltonian $H$ as:
\begin{equation}
H = M^{-1}H_0.
\label{eq:hamiltonian}
\end{equation}
Clearly, this new system can have real eigenvalues $E_n$. However, the Hamiltonian $H$ in Eq.~\ref{eq:hamiltonian} is generally non-Hermitian (\textit{i.e}, $H\neq H^\dagger$). 
To understand this non-Hermiticity, after performing the  Hermitian operation on $H$, one can obtain that $H^\dagger = (M^{-1}H_0)^\dagger = H_0M^{-1}$ and usually $H^\dagger = H_0M^{-1} \neq M^{-1}H_0 = H$, as a result of  the non-commutativity for the matrix production, except for some special cases such as when $H_0$ is diagonal or the components in $M$ are uniformly identical, namely $M = c \mathbb{I}$, where $\mathbb{I}$ is the identity matrix and $c$ is the arbitrary constant.

Although $H$ in Eq.~\ref{eq:hamiltonian} is non-Hermitian, it can have real eigenvalues, which means that there is no point gap in the bandstructure but only one special line gap that is perpendicular to the real axis in complex energy plane. 
Based on this special line gap, it is possible for the non-Hermitian Hamiltonian of Eq.~\ref{eq:hamiltonian} to possess nontrivial topology~\cite{Bergholtz2021, Kawabata2019a}. Interestingly, the topology can be either inherited from a topologically non-trivial $H_0$, or induced by the $M$ matrix with a trivial $H_0$. 

\begin{figure}[tp!]
\centering
\includegraphics[width=\linewidth]{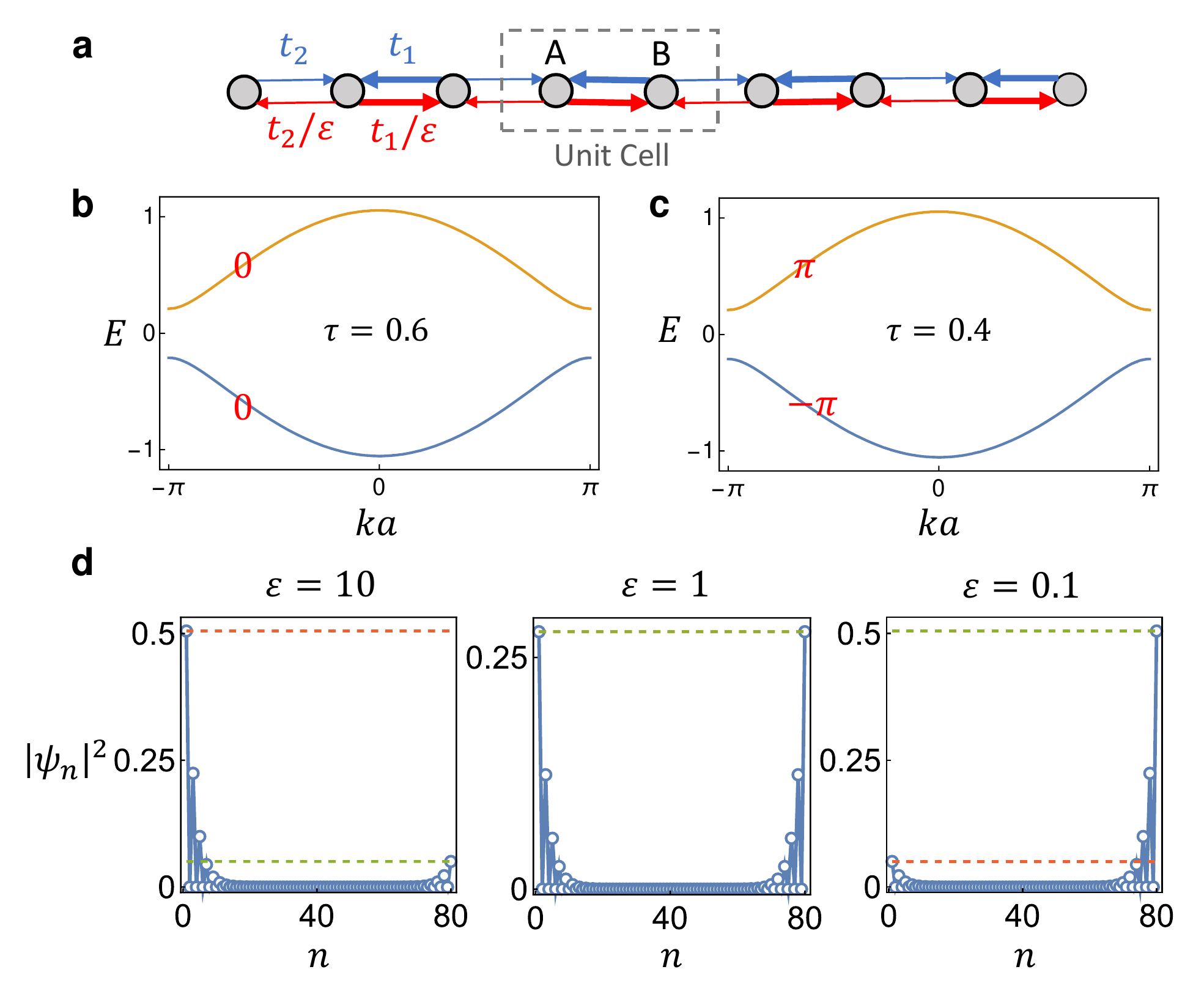}
\caption{1D non-Hermitian topological system with real spectra. (a) System settings. $t_1 = \kappa_0 \tau$ and $t_2= \kappa_0 (1-\tau)$, $\kappa_0 = 1$, $\tau \in (0,1)$, $\varepsilon \in \mathbb{R}$. (b,c) Bandstructures when $\tau=0.6$ and $\tau = 0.4$. Here, $\varepsilon = 0.9$. The red number on each band denotes the Zak phase. (d) The topological edge states will localize toward one end more than the other when $\varepsilon \neq 1$. The system has 80 sites ($N=80$). The $\psi_n$ denotes the field on the $n$-th site. The blue line with circle markers represents the field distribution. The red and green dotted lines denote the field amplitudes at the left and right ends, respectively. }
\label{fig:1DnonHermitian}
\end{figure}

\textit{1D non-Hermitian system from the SSH model.} As the first example, we discuss a 1D non-Hermitian system with coupling settings shown in Fig.~\ref{fig:1DnonHermitian}(a), similar to the conventional SSH model~\cite{Su1979}. The corresponding Hamiltonian is:
\begin{equation}
\begin{aligned}
H &= \sum_n t_1 a^\dagger_n b_n + \frac{t_1}{\varepsilon} b^\dagger_n a_n + t_2 a^\dagger_{n+1} b_n + \frac{t_2}{\varepsilon} b^\dagger_{n} a_{n-1}, 
\label{eq:1DH}
\end{aligned}
\end{equation}
where $t_1 = \kappa_0 \tau$, $t_2 = \kappa_0 (1-\tau)$, $\tau \in (0,1)$, and $a^\dagger$($a$) and $b^\dagger$($b$) are the creation (annihilation) operators. It can be regarded as the production of the inverse matrix of $M = {\rm diag} [1,\varepsilon,1, \varepsilon,...] $ and the Hamiltonian of the conventional SSH model $H_0 = \sum_n \left( t_1 b^\dagger_n a_n + t_2 b^\dagger_{n} a_{n-1} + C.C \right) $, namely, $H = M^{-1}H_0$. The Hamiltonian of Eq.~\ref{eq:1DH} in the momentum space can be expressed as:
\begin{equation}
\begin{aligned}
H_k &= (t_1 + t_2 \cos(ka)) \left(\frac{1+\varepsilon}{2\varepsilon}\sigma_x + i \frac{\varepsilon-1}{2\varepsilon} \sigma_y\right) \\
&+ t_2 \sin(ka)  \left(\frac{1+\varepsilon}{2\varepsilon}\sigma_y - i \frac{\varepsilon-1}{2\varepsilon} \sigma_x\right),
\end{aligned}
\label{eq:1DHk}
\end{equation}
where $k$ is the Bloch wave vector, $a$ is the lattice constant and $\sigma_i$ are the Pauli matrices. This Hamiltonian satisfies the chiral symmetry: $H_k = -\sigma_z H_k \sigma_z$. We can easily obtain its energy spectra: $E_k = \pm \frac{1}{\sqrt{\varepsilon}}\sqrt{t_1^2 + t_2^2 + 2t_1 t_2 \cos(k)}$, as plotted in Fig.~\ref{fig:1DnonHermitian}(b,c). The spectra are always real even for strong non-Hermiticity (i.e, $\varepsilon \gg 1$ or $\varepsilon \ll 1$). 
Note that some other non-Hermitian systems based on non-reciprocal couplings, e.g., the Hatano-Nelson model~\cite{Hatano1996}, can exhibit real energy spectra under the open boundary condition, but these real spectra are a result of boundary effect, being extremely sensitive to boundary conditions~\cite{Hatano1997, Hatano1998, Kunst2018, Lee2019a, Koch2020, Xiong2018}. In contrast, the real spectra of our systems are insensitive to boundary conditions (see detailed discussions in Supplementary Material~\cite{Note1}).

\begin{figure}[tp!]
\centering
\includegraphics[width=\linewidth]{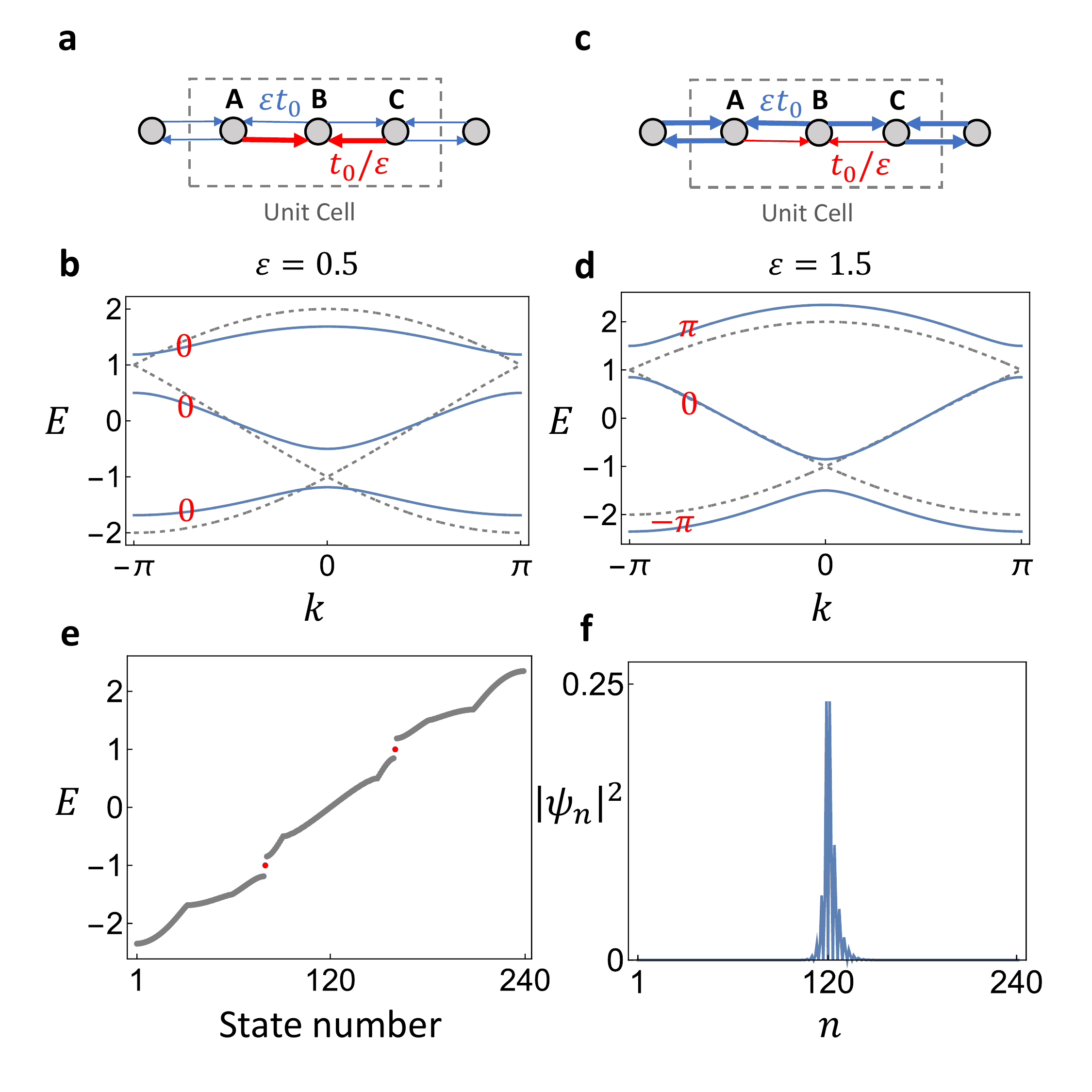}
\caption{Non-Hermiticity-induced 1D topological phase transition. (a) The unit cell when $\varepsilon<1$. The direction and thickness of the arrow represent the coupling direction and strength, respectively. (b) The bandstructure of non-Hermitian system ($\varepsilon=0.5$) with trivial band topology. The red number on each band labels the Zak phase. The blue solid and gray dotted lines denote the bandstructures with $\varepsilon=0.5$ and $\varepsilon=1$, respectively. (c) The unit cell when $\varepsilon>1$. (d) The bandstructure of non-Hermitian system ($\varepsilon=1.5$) with non-trivial band topology. (e) The topological interface states exist in the bulk gap of the junction system composed of two topologically distinct lattices ($\varepsilon=0.5$ and $\varepsilon=1.5$). Each lattice has 120 sites. The interface and bulk states are represented by the red and gray points, respectively. (f)  The field distribution of the topological interface state at $E=1$. The $\psi_n$ denotes the field on the $n$-th site. Here, $t_0=1$.}
\label{fig:1DnonHermitian_epsilon}
\end{figure}

The topology of our 1D non-Hermitian system can be characterized by the Zak phase: $c_n = -i\int_{-\pi/a}^{\pi/a} \langle w_n| \partial_k |u_n\rangle dk$, where $H_k|u_{k,n}\rangle = E_{k,n} |u_{k,n}\rangle$ and $H^{\dagger}_k |w_{k,n} \rangle= E_{k,n} |w_{k,n}\rangle$. 
The Zak phases  $(c_1, c_2)$ for the two bands are: $(c_1, c_2) = (0,0)$ for $t_1 > t_2$, and $(c_1, c_2) = (-\pi,\pi)$ for $t_1<t_2$~\cite{Zak1989, Note1}. 
The non-zero Zak phase for $t_1<t_2$ indicate that there are two degenerate topological edge states localized at the ends of a finite system, as a manifestation of fractional charges at the ends~\cite{Vanderbilt1993, Resta1994, Bradlyn2017, Benalcazar2019, Benalcazar2017}. 
From Eq.~\ref{eq:1DH}, we can see that the mirror symmetry is broken when $\varepsilon \neq 1$. 
The topological states can be affected by this non-Hermitian parameter. 
For a finite system with $N$ sites, the ratio between the amplitudes of the left and right edge states is $\varepsilon$~\cite{Note1}: $\frac{|\psi_{1}|^2}{|\psi_{N}|^2} = \varepsilon $, where $\psi_{1/N}$ denote the field on the left/right end site.
As shown in Fig.~\ref{fig:1DnonHermitian}(d), when $\varepsilon \neq 1$, the topological edge states can localize toward one end more than the other~\cite{Note1}.

\textit{Non-Hermiticity-induced topological phase transition in 1D system.}
After demonstrating the topology inherited from the Hermitian topological system $H_0$ in Eq.~\ref{eq:1DH}, we now show that the topology can also be induced by the non-Hermitian parameter $\varepsilon$ that is applied to an originally trivial 1D $H_0$.
As shown in Fig.~\ref{fig:1DnonHermitian_epsilon}(a,c), the Hamiltonian is:
\begin{equation}
\begin{aligned}
H = t_0 \sum_n & \frac{1}{\varepsilon} b_n^\dagger a_n + \varepsilon a_n^\dagger b_n + \frac{1}{\varepsilon} b_n^\dagger c_n + \varepsilon c_n^\dagger b_n \\
&+ \varepsilon (a_{n+1}^\dagger c_n + c_n^\dagger a_{n+1}),
\end{aligned}
\label{eq:1DHkespilon}
\end{equation}
which is constructed by a trivial Hermitian system $H_0 = t_0 \sum_n b_n^\dagger a_n + c_n^\dagger b_n + a_{n+1}^\dagger c_n + C.C.$ and $M = {\rm diag}[\varepsilon^{-1}, \varepsilon, \varepsilon^{-1},\varepsilon^{-1}, \varepsilon, \varepsilon^{-1}...]$, namely, $H=M^{-1}H_0$.
The Hamiltonian of Eq.~\ref{eq:1DHkespilon} in the momentum space can be written as:
\begin{equation}
H_k = t_0 \begin{pmatrix}
0 & \varepsilon & \varepsilon e^{-i ka} \\
\frac{1}{\varepsilon} & 0 & \frac{1}{\varepsilon} \\
\varepsilon e^{i ka} & \varepsilon & 0
\end{pmatrix},
\label{eq:Hk1DnonHermitian}
\end{equation}
where $k$ is the wave vector, $a$ is the lattice constant.
Obviously, when $\varepsilon = 1$, the $H_k$ represents a gapless Hermitian system. 
However, after setting $\varepsilon \neq 1$, the spectra of $H_k$ of Eq.~\ref{eq:Hk1DnonHermitian} can open a gap as presented in Fig.~\ref{fig:1DnonHermitian_epsilon}(b,d). 

Importantly, when $\varepsilon <1$ ($\varepsilon>1$), the system becomes trivial (topological). 
Compared with the system with $\varepsilon<1$ in Fig.~\ref{fig:1DnonHermitian_epsilon}(b), the system with $\varepsilon>1$, as shown in Fig.~\ref{fig:1DnonHermitian_epsilon}(d), can lead to the non-zero Zak phases for the first and third bands. 
After combining the trivial and topological systems into a chain, topological interface states emerge in the bulk gaps, as shown in Fig.~\ref{fig:1DnonHermitian_epsilon}(e,f). 
This shows that the non-Hermitian parameter $\varepsilon$ not only makes the $H$ of Eq.~\ref{eq:hamiltonian} become non-Hermitian, but is also able to drive a topological phase transition. 

\begin{figure}[tp!]
\centering
\includegraphics[width=\linewidth]{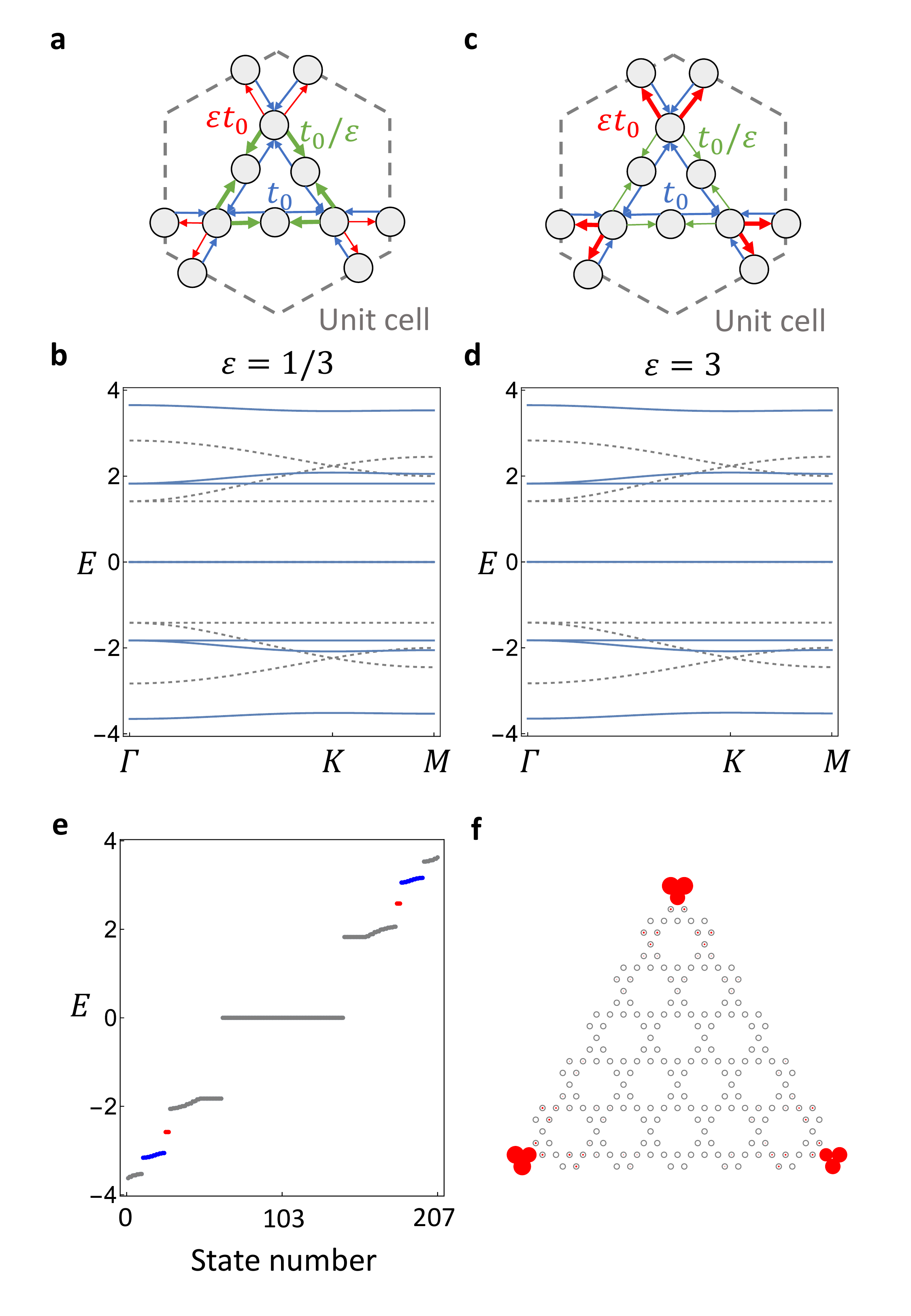}
\caption{Non-Hermiticity-induced higher-order topological states in  2D $C_3$-symmetric lattice. (a) The unit cell when $\varepsilon<1$, which contains 9 sites. (b) The bandstructure of the non-Hermitian system with $\varepsilon=1/3$. The blue solid and gray dotted lines denote the bandstructures with $\varepsilon=1/3$ and $\varepsilon=1$, respectively. (c) The unit cell when $\varepsilon>1$. (d) The bandstructure of the non-Hermitian system with $\varepsilon=3$. (e) The eigenvalues for a finite triangular system in (f) with $\varepsilon=3$. The red, blue and gray points denote the corner, edge and bulk states, respectively. (f) The field distribution of topological corner states at $E=-2.58$. The size of red points indicates the field amplitude on the sites. Here, $t_0 = 1$.}
\label{fig:2DnonHermitianHOTI}
\end{figure}

\textit{Non-Hermiticity-induced higher-order topological states.}
We then proceed to demonstrate the non-Hermiticity-induced higher-order topological states in a 2D lattice with $C_3$ symmetry.
The unit cell of the system is shown in Fig.~\ref{fig:2DnonHermitianHOTI}(a,c). 
When $\varepsilon = 1$, the system corresponds to a trivial $C_3$-symmetric Hermitian system.
The condition of $\varepsilon \neq 1$ can not only make the system non-Hermitian, but also open a gap at the valley point $K$, as illustrated in Fig.~\ref{fig:2DnonHermitianHOTI}(b,d)~\cite{Note1}.  
Notably, the systems with $\varepsilon<1$ and $\varepsilon>1$ correspond to trivial and topological ones, respectively.

The higher-order topological properties of the $C_3$-symmetric non-Hermitian systems can be described by the  topological index and symmetry indicator~\cite{Benalcazar2019, Liu2021a, Okugawa2021}.
The topological index is given by $\chi = ([K_1^{(3)}], [K_2^{(3)}])$, where $[K_1^{(3)}]=\sharp K_1^{(3)} - \sharp \Gamma_1^{(3)}$, $[K_2^{(3)}]=\sharp K_2^{(3)} - \sharp \Gamma_2^{(3)}$ are the symmetry indicators. 
The $\sharp \Pi_p^{(n)}$ is the number of bands below the bandgap with the eigenvalue of $C_3$ operator $e^{i2\pi (p-1)/n}$ ($p=1,2,...,n$) at the high-symmetry point $\Pi=\Gamma, K$. 
We find that $\chi = (-1,1)$ when $\varepsilon>1$ and $\chi=(0,0)$ when $\varepsilon<1$. 
Remarkably, the topological bulk polarization $\bm{P}$ and fractional corner charge $Q$ can be determined by the topological index~\cite{Benalcazar2019}: $\bm{P}=  \left(\frac{2}{3}[K_1^{(3)}] + \frac{4}{3}[K_2^{(3)}] \right)
 (\bm{a}_1 + \bm{a}_2) \mod 1$, where $\bm{a}_{1,2}$ are primitive unit lattice vectors, $Q = \frac{1}{3} [K_2^{(3)}]\mod 1$. When $\varepsilon<1$, $\bm{P}=\bm{0}$ and $Q=0$. When $\varepsilon>1$, $\bm{P}=\frac{2}{3}(\bm{a}_1 + \bm{a}_2)$ and $Q=1/3$. 
Due to the non-zero $\bm{P}$ and $Q$ when $\varepsilon>1$, we can see that there are corner states in the bulk gap for a finite system, as shown in Fig.~\ref{fig:2DnonHermitianHOTI}(e, f).

\begin{figure}[tp!]
\centering
\includegraphics[width=\linewidth]{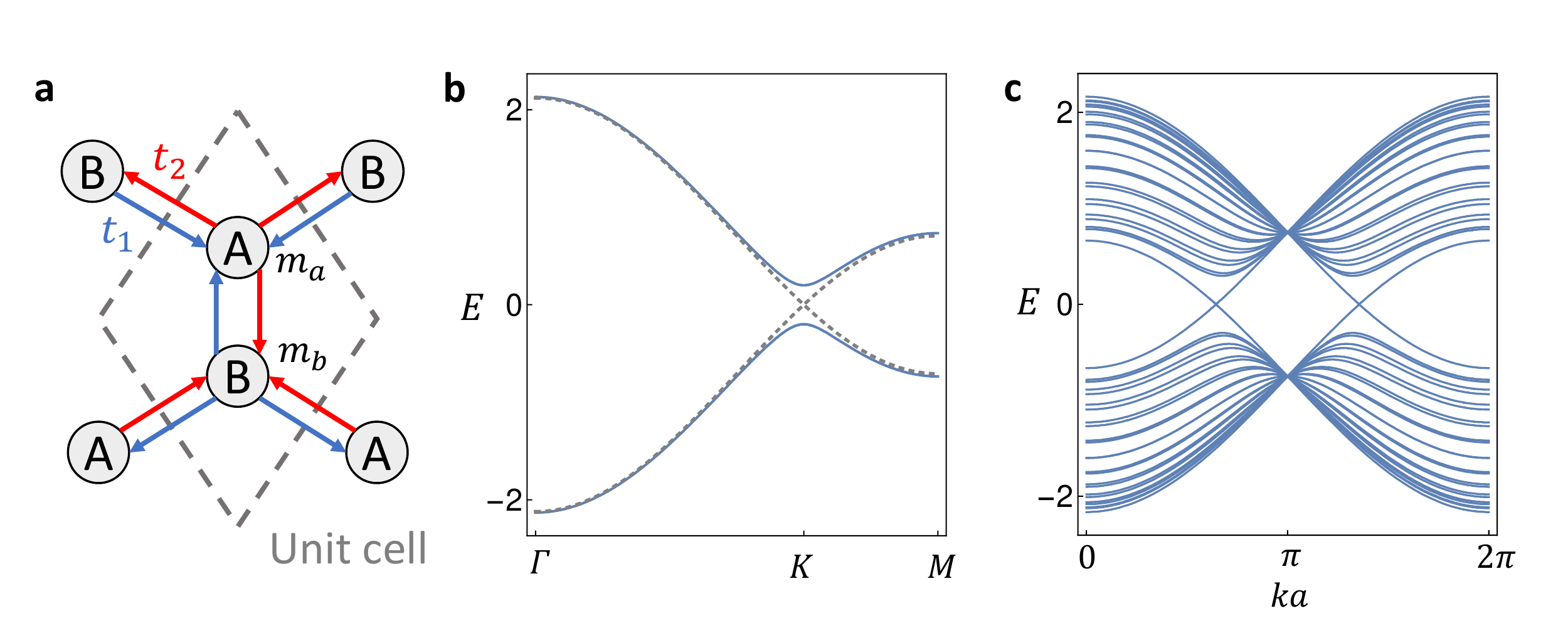}
\caption{Non-Hermitian topological valley system. (a) The unit cell of the 2D honeycomb lattice. The $t_1$ and $t_2$ are the coupling strengths, $t_2 =\frac{t_1}{\varepsilon}$, $\varepsilon \neq 1$. (b) The bandstructure of the non-Hermitian system with $\varepsilon=2.0$. Here, $t_1 = -1.0$, $m_a = -m_b = m$. The bandstructures with $m=0.2$ and $m=0$ are represented by the blue solid and gray dashed lines, respectively. (c) The dispersion of the topological edge states on the zigzag interface between two non-Hermitian systems with $m=0.25$ and $m=-0.25$.}
\label{fig:2DnonHermitianvalley}
\end{figure}

\textit{Non-Hermitian effects on topological edge states in a 2D honeycomb lattice.} Here, we discuss the non-Hermitian topological valley system in 2D honeycomb lattice, which has the unit cell in Fig.~\ref{fig:2DnonHermitianvalley}(a). 
The effective non-Hermitian Hamiltonian at the valley point $K$ ($\bm{k} \rightarrow \bm{k}_0 + \delta \bm{k}$, $\bm{k}_0$ is the wave vector of $K$) is~\cite{Note1}:
\begin{equation}
\begin{aligned}
H_k &= v_F \delta k_x \left(\frac{1+\varepsilon}{2\varepsilon}\sigma_x +  i \frac{\varepsilon-1}{2\varepsilon} \sigma_y \right) \\
&+ v_F \delta k_y \left(\frac{1+\varepsilon}{2\varepsilon}\sigma_y -  i \frac{\varepsilon-1}{2\varepsilon} \sigma_x\right) + m \sigma_z,
\label{eq:2DHkvalley}
\end{aligned}
\end{equation}
where $\varepsilon \neq 1$, $m_a = -m_b = m$ and $v_F = - \frac{3}{2}t_1 a$, $a$ is the lattice constant. 
Setting $m \neq 0$ can open a gap at the valley point $K/K'$, as shown in Fig.~\ref{fig:2DnonHermitianvalley}(b). 
The topological properties of this non-Hermitian valley system can be described by the valley Chern number of the lower band: $C_v =- \frac{{\rm sign}[m]}{2}$~\cite{Note1, Lu2016}. 
After combining two lattices with different valley Chern numbers ($m>0$ and $m<0$) along the zigzag boundary, there are topological edge states existing in the bulk gap ~\cite{Wang2018}, as shown in Fig.~\ref{fig:2DnonHermitianvalley}(c). 
These edge states have the dispersion relation ~\cite{Note1}: $E = \pm \frac{v_F}{\sqrt{\varepsilon}} k $, where $k$ is the wave vector of an edge state.
This dispersion shows that the $\varepsilon$ can affect the propagation velocity of edge states $|v_{\rm edge}| = |\partial_k E| = \frac{|v_F|}{\sqrt{\varepsilon}}$. 
When $\varepsilon < 1$ ($\varepsilon > 1$), the propagation velocity of edge states can become faster (slower) than in the Hermitian case: $|v_{\rm edge}|>|v_F|$ ($|v_{\rm edge}|<|v_F|$).

\textit{Experimental proposals.}  The non-Hermitian systems shown above can be achieved by electric circuits~\cite{Albert2015, Imhof2018,Zhu2019,Zhu2018,Liu2021b}. The electric diode, which makes the electric current flow directionally, can be used to readily establish non-reciprocal coupling~\cite{Ezawa2019}. As an example, we demonstrate the circuit realization of 1D non-Hermitian topological system in Ref.~\footnote{See supplementary materials for more details about the Hamiltonians, the non-sensitivity of our systems to boundary conditions, the effect of the non-diagonal $M$, topologically valley-protected edge states and experimental proposals, which include the Ref.~\cite{Ni2018, Zeng2021, Brody2013, Xiao2019, Kawabata2017,Regensburger2012, Ezawa2018, Ezawa2019a, Ningyuan2015}}. 
There are many approaches for classical wave systems to achieve non-reciprocal coupling. 
For example, the non-reciprocal coupling in optical wave systems can be achieved by exploiting dynamically modulated media~\cite{Yu2009}, metamaterials~\cite{Sounas2013} or S-bending waveguides~\cite{Liu2021}. 
For acoustic or elastic wave systems, one can achieve non-reciprocal coupling by using piezophononic media~\cite{Gao2020}, metamaterials~\cite{Liang2010} or additional electric setup~\cite{zhang2021acoustic}. 

To summarize, we propose a systematic strategy to construct real-eigenvalued non-Hermitian topological systems. 
Compared with previous non-Hermitian systems whose real spectra exist only when non-Hermiticity is relatively weak, our systems exhibit energy spectra that are always real, regardless of weak or strong non-Hermiticity. 
Furthermore, the nonreciprocal-coupling-based non-Hermiticity is able to affect many properties of topological states. For example, it is able to determine the topology by inducing a topological phase transition. 
Our work offers a new avenue to explore non-Hermitian topological physics~\cite{Zeng2020, Liu2019,Lee2019, Xiong2021, zeng2021real}, and would be useful in novel photonic and acoustic devices with non-reciprocal coupling.

\begin{acknowledgments}
This research is supported by Singapore National Research Fundation Competitive Research Program under Grant no. NRF-CRP23-2019-0007, and Singapore Ministry of Education Academic Research Fund Tier 3 under Grant No. MOE2016-T3-1-006 and Tier 2 under Grant No. MOE2018-T2-1-022 and Grant No. MOE2019-T2-2-085.
\end{acknowledgments}

\bibliography{references}

\end{document}